\documentclass[prl,aps,showpacs,twocolumn,preprintnumbers,
amsmath,amssymb,superscriptaddress,longbibliography]{revtex4-1}
\usepackage[english]{babel}
\usepackage{amsmath,amssymb,amsfonts}
\usepackage{graphicx}
\usepackage[colorlinks=True,linkcolor=red,citecolor=blue,urlcolor=blue]{hyperref}
\usepackage{bookmark}
\usepackage[dvipsnames]{xcolor} 
\usepackage{braket}
\usepackage{nicefrac}
\usepackage{bm}
\usepackage{bbm}
\usepackage{braket}
\usepackage{slashed}
\usepackage[normalem]{ulem}
\usepackage{array}
\usepackage{makecell}
\newcolumntype{C}{>{$}c<{$}}

\def\bk{\mathbf{k}}
\def\id{\mathbbm{1}}

\newcommand{\tr}{\mathop{\mathrm{tr}}}

\renewcommand{\Re}{\mathop{\mathrm{Re}}}
\renewcommand{\Im}{\mathop{\mathrm{Im}}}
\newcommand{\HH}{\mathcal{H}}

\newcommand{\bp}{\mathbf{p}}

\newcommand{\Z}{\mathbb{R}}
\renewcommand{\i}{\mathrm{i}}
\renewcommand{\HH}{\mathcal{H}}

\newcommand{\ave}[1]{\langle #1 \rangle}
\newcommand{\bolds}[1]{\boldsymbol #1}

\allowdisplaybreaks
\DeclareMathAlphabet{\zc}{OT1}{pzc}{m}{it}

\def\ket#1{|#1\rangle }
\def\bra#1{\langle #1 |}

  \DeclareUnicodeCharacter{2212}{-}

\begin{document}

\title{Protection of all nondefective twofold degeneracies\\ by anti-unitary symmetries in non-Hermitian systems}
\author{Sharareh Sayyad}
\email{sharareh.sayyad@mpl.mpg.de}
\affiliation{Max Planck Institute for the Science of Light, Staudtstra\ss e 2, 91058 Erlangen, Germany}
\date{\today}

\begin{abstract}
    Non-Hermitian degeneracies are classified as defective exceptional points~(EPs) and nondefective degeneracies. While in defective EPs, both eigenvalues and eigenvectors coalesce, nondefective degeneracies are characterized merely by the emergence of degenerate eigenvalues. It is also known that all degeneracies are either symmetry-protected or accidental. In this paper, I prove that anti-unitary symmetries protect all nondefective twofold degeneracies. By developing a 2D non-Hermitian tight-binding model, I have demonstrated that these symmetries comprise various symmetry operations, such as discrete or spatial point-group symmetries and Wick's rotation in the non-Hermitian parameter space. Introducing these composite symmetries, I present the protection of nondefective degeneracies in various parameter regimes of my model. This work paves the way to stabilizing nondefective degeneracies and offers a new perspective on understanding non-Hermitian band crossings.

\end{abstract}

\maketitle

\paragraph{\bf Introduction.---}
Appearance of degeneracies in the energy spectra of different Hermitian systems gives rise to a plethora of phenomena such as quantized classical~\cite{Imhof2018,  Kotwal2021, Li2021} and quantum~\cite{Miao2014, DeJuan2017, Tran2017, Hubener2017, Armitage2018, FloresCalderon2021} responses, quantum anomalies~\cite{Gooth2017, Das2020, Ong2021,  Zeng2022b, Zeng2022} and the emergence of novel effective quasiparticles~\cite{Bradlyn2016, Tang2017, Ma2018, Cano2019, Lv2019, Xie2021}. The occurrence of degenerate energy levels historically has been classified into either symmetry-protected or accidental degeneracies~\cite{Herring1937, vonNeumann1993,  Demkov2007, Xu2016, Allen2018}. Here, accidental degeneracies refer to the intersection of energy levels due to the fine-tuning of parameters without symmetry stabilization. It has been later discussed that all band-touching points in two-band Hermitian systems are protected by anti-unitary symmetries dubbed "hidden symmetries"~\cite{Hou2013, Hou2017, Hou2018, Hou2020}. Accidental degeneracies are stable in these systems as long as these hidden symmetries are respected. It has further been shown that these hidden symmetries are usually composite of various discrete operations, including rotation, translation, sublattice exchange, and complex conjugation.

Another platform where degeneracies play a crucial role is in non-Hermitian physics, which effectively describes open systems. This field of study encounters surges of interest as some of its underlying properties have no Hermitian counterparts~\cite{Ashida2020, Bergholtz2021, Okuma2022}. The appearance of defective exceptional points~(EPs)~\cite{ Lin2019, Fu2022}, at which both eigenvalues and eigenvectors coalesce, and the accumulation of bulk modes on the boundaries, known as the skin effect~\cite{Borgnia2020, Okuma2020, Kawabata2020, Okuma2021, Zhang2022, Longhi2022}, exemplify prominent properties of non-Hermitian systems which cannot be realized in Hermitian setups. Aside from these possibilities, non-Hermitian systems may accommodate other degeneracies which are nondefective~\cite{Shen2018, Xue2020, Yang2021, Sayyad2022b, Wiersig2022}. These nondefective degeneracies can be further classified into two classes. While one type of these nondefective degeneracies has an analog in Hermitian physics and is usually isolated~\footnote{ This type of degeneracies are sometimes dubbed 'diabolic points' or 'nodal points'~\cite{Shen2018, Sayyad2022b}.}, the other type resides in the vicinity of defective EPs~\footnote{These nondefective degeneracies are also known as the 'nondefective EPs'~\cite{Sayyad2022b}.} and hence has no counterparts in Hermitian physics~\cite{Sayyad2022b}. All of these non-Hermitian degeneracies, as well as the skin effect in non-Hermitian systems, is under theoretical investigation and experimental observation in various field of research, including classical active matters~\cite{Sone2020, Palacios2021}, classical electric circuits~\cite{Ezawa2019, Hofmann2020, RafiUlIslam2021, Wu2022}, quantum circuits~\cite{Fleckenstein2022}, photonics~\cite{Valagiannopoulos2018, Wang2021, Parto2021, Jin2022, Valagiannopoulos2022}, phononics~\cite{Liu2022, DelPino2022}, laser physics~\cite{Peng2014, Feng2014, Hodaei2015}, field theories~\cite{Alexandre2018, Sayyad2021, Kawabata2021}, transport physics~\cite{Longhi2017, Du2020, Ghaemi-Dizicheh2021, Franca2022} and non-equilibrium dynamics~\cite{Sayayd2021, Zhai2022, Starchl2022}.

The spate of studies on EPs identifies numerous (spatial or discrete) symmetries which protect defective EPs~\cite{Yoshida2019, Delplace2021, Mandal2021, Sayyad2022, Yoshida2022, Sayyad2022b, Cui2022}. It has also been shown that the intersection of an even number of symmetry-protected higher-dimensional defective EPs, e.g., exceptional rings, results in observing nondefective degeneracies~\cite{Kirillov2013, Sayyad2022b, Cui2022}.
Hence, all nondefective degeneracies found in these situations are also stabilized by symmetry, which protects the defective EPs~\cite{Yoshida2019, Yoshida2019b, Mandal2021}.
Further attempt regarding symmetry stabilization of isolated nondefective degeneracies is based on a case study on four-band models and in the presence of two symmetries, namely pseudo-Hermiticity and anti-parity-time symmetries, which impose strict restrictions on the eigenspace of the model~\cite{Xue2020}. However, the analog between well-studied Hermitian degeneracies and nondefective degeneracies urges one to go beyond the case studies and identify the key factors that make any nondefective degeneracies robust. 

As I have pointed out, all accidental degeneracies in two-band Hermitian systems are symmetry protected. One may wonder whether the stabilization of Hermitian accidental degeneracies can be extended to the realm of non-Hermitian physics. In this Letter, I prove that composite anti-unitary symmetries protect all twofold nondefective degeneracies in non-Hermitian models. I further demonstrate that, due to the biorthogonality of the eigenspace, these symmetry operators come in (right and left) pairs. By introducing a 2D non-Hermitian tight-binding model, I show that different non-Hermitian composite symmetries protect nondefective degeneracies in various parameter regimes in my model. 
These composite symmetries are distinct from their Hermitian counterparts due to the presence of a Wick's rotation in the non-Hermitian parameter regime.

\paragraph{\bf Theorem.---}
In the following, I prove that {\it all nondefective twofold degeneracies in non-Hermitian systems are protected by anti-unitary operations with nonunity square.}

The eigensystem of a two-band non-Hermitian Hamiltonian $\HH_{\rm nH} $ with nondefective degeneracy~($\lambda=\lambda_{0}$) casts
\begin{align}
\HH_{\rm nH} \ket{\psi^{R}_{i}}= \lambda_{0} \ket{\psi^{R}_{i}}, \quad & \quad \bra{\psi^{L}_{i}} \HH_{\rm nH} = \bra{\psi^{L}_{i}} \lambda_{0}
\label{eq:nH1}
,\\
\HH^{\dagger}_{\rm nH} \ket{\psi^{L}_{i}}= \lambda_{0}^{*} \ket{\psi^{L}_{i}}, \quad & \quad  \bra{\psi^{R}_{i}} \HH^{\dagger}_{\rm nH} = \bra{\psi^{R}_{i}} \lambda_{0}^{*}.
\label{eq:nH2}
\end{align}
where $\ket{\psi^{R/L}_{i}}$ with $i \in \{1,2\}$ denotes the right/left biorthogonal eigenvector such that $\ave{ \psi^{L}_{j} |\psi^{R}_{i}} = \delta_{ij}$.

In Hermitian systems, the Hermiticity imposes that vanishing the commutation relation between the symmetry operator and the Hamiltonian leads to the invariance of the system under a particular symmetry. However, due to the lack of Hermiticity in non-Hermitian models, biorthogonality of eigenvectors necessitates introducing left and right symmetry operations such that, see also the Supplemental Material~(SM)~\cite{SuppMat},
\begin{align}
  & \HH_{\rm nH}^{\dagger} \Upsilon^{R}_{\rm nH}- \Upsilon^{R}_{\rm nH} \HH_{\rm nH} =0,\label{eq:symnhR}
  \\
   & \HH_{\rm nH} \Upsilon^{L}_{\rm nH}- \Upsilon^{L}_{\rm nH} \HH^{\dagger}_{\rm nH}  =0 .\label{eq:symnhL}
\end{align}
 The explicit form of these symmetry operators in terms of biorthogonal basis reads~\footnote{We note that Hermitian-like nondefective degeneracies exhibit no algebraic singularities and possess stable Jordan normal form. However, the Jordan decomposition for the other nondefective degeneracies, located close to defective EPs, is unstable~\cite{Sayyad2022b}. These subtleties in eigenvectors of two types of nondefective degeneracies should be reflected in symmetry operators~($\Upsilon^{R}_{\rm nH},\Upsilon^{L}_{\rm nH}$). }
\begin{align}
        \Upsilon^{R}_{\rm nH} &= \left[ \ket{\psi_1^{R}} \bra{\psi_2^{R*}} - \ket{\psi_2^{R}} \bra{\psi_1^{R*}} \right] {\cal K},
    \label{eq:finalUpsRnh}
    \\
    \Upsilon^{L}_{\rm nH} &=  \left[ \ket{\psi_1^{L}} \bra{\psi_2^{L*}} - \ket{\psi_2^{L}} \bra{\psi_1^{L*}} \right] {\cal K},
    \label{eq:finalUpsLnh}
\end{align}
where $\cal K$ is the complex conjugation operator that ensures $\Upsilon^{R/L}_{\rm nH} $ is anti-unitary. Employing Eqs.~(\ref{eq:finalUpsRnh}, \ref{eq:finalUpsLnh}), one can verify that $\Upsilon^{R}_{\rm nH} \cdot \Upsilon^{L}_{\rm nH}=\Upsilon^{L}_{\rm nH}  \cdot \Upsilon^{R}_{\rm nH}=-\id$\footnote{the centerdot~('$\cdot$') represents the standard composition product.}.  Applying these symmetries on biorthogonal eigenvectors then yields
\begin{align}
    \Upsilon^{R}_{\rm nH} \ket{\psi^{L}_{2}}&= \ket{\psi_{1}^{R}},
    \quad 
      \Upsilon^{R}_{\rm nH} \ket{\psi^{L}_{1}}= -\ket{\psi_{2}^{R}},
      \label{eq:URpsi}
    \\
    \Upsilon^{L}_{\rm nH} \ket{\psi^{R}_{2}}&= \ket{\psi_{1}^{L}}, 
    \quad 
        \Upsilon^{L}_{\rm nH} \ket{\psi^{R}_{1}}= -\ket{\psi_{2}^{L}}.
         \label{eq:ULpsi}
\end{align}
From Eqs.~(\ref{eq:URpsi}, \ref{eq:ULpsi}), I realize that both eigenvectors in $
(\Upsilon^{R}_{\rm nH} \ket{\psi^{L}_{1}},\ket{\psi_1^{R}} )
$  or in $(\Upsilon^{L}_{\rm nH} \ket{\psi^{R}_{1}}, \ket{\psi_1^{L}})$ possess the degenerate eigenvalue $\lambda_{0}$ or $\lambda^{*}_{0}$, respectively.

To corroborate that these degeneracies are protected by symmetry with a nonunity square, I in the following show that eigenvectors are orthogonal by calculating the overlap of eigenvectors at the nondefective degenerate point. 
For an anti-unitary operator $\Upsilon$ in Hermitian systems, one can rewrite the overlap of eigenvectors as $\ave{\psi | \phi}=\ave{\Upsilon \phi | \Upsilon \psi}$. In non-Hermitian setups, biorthogonality of eigenvectors results in generalizing this relation using $\Upsilon^{R}$ and $\Upsilon^{L}$ into
\begin{align}
\begin{cases}
\ave{\psi^{L} | \phi^{R}}&=\ave{\Upsilon^{L} \phi^{R} | \Upsilon^{R} \psi^{L}},\\
\ave{\psi^{R} | \phi^{L}}&=\ave{\Upsilon^{R} \psi^{L} | \Upsilon^{L} \phi^{R}}.
\end{cases}
\end{align}

Having this and by defining $\Upsilon^{L}_{\rm nH} \ket{\psi_{1}^{R} } =\ket{ \tilde{ \psi}_{1}^{L}} $ and $\Upsilon^{R}_{\rm nH} \ket{\psi_{1}^{L} } =\ket{ \tilde{ \psi}_{1}^{R}}$, I evaluate the inner products as
\begin{align}
    \ave{\psi_{1}^{L} | \tilde{\psi}_{1}^{R}} &=
    \ave{\Upsilon^{L}_{\rm nH} \tilde{\psi}^{R}_{1}  | \Upsilon_{\rm nH}^{R} \psi^{L}_{1} }
,\nonumber \\
&=\ave{  
\Upsilon_{\rm nH}^{L} \Upsilon_{\rm nH}^{R}\psi^{L}_{1} 
|
\Upsilon_{\rm nH}^{R} \psi_{1}^{L}
},\nonumber \\
&=
- \ave{
\psi_{1}^{L}
| \Upsilon_{\rm nH}^{R} \psi^{L}_{1}
},\nonumber \\
    & =
     -\ave{\psi^{L}_{1} |\tilde{\psi}_{1}^{R}  }
     \label{eq:LtR}
, \\
    \ave{\psi_{1}^{R} | \tilde{\psi}_{1}^{L}} &=
\ave{\Upsilon_{\rm nH}^{R} \tilde{\psi}^{L}_{1} | \Upsilon_{\rm nH}^{L} \psi^{R}_{1}  }
,\nonumber \\
&=
\ave{ 
\Upsilon_{\rm nH}^{R} \Upsilon_{\rm nH}^{L} \psi_{1}^{R}
|
\Upsilon_{\rm nH}^{L} \psi_{1}^{R}
}
,\nonumber \\
&=- \ave{\psi^{R}_{1} | \Upsilon_{\rm nH}^{L} \psi_{1}^{R}}
,\nonumber \\
    & =
     -\ave{\psi^{R}_{1} |\tilde{\psi}_{1}^{L}  }.
   \label{eq:tLR}
\end{align}
The above relations can be also verified using Eq.~(\ref{eq:URpsi}, \ref{eq:ULpsi}).
Here I set $ \Upsilon^{R}_{\rm nH} \Upsilon^{L}_{\rm nH} \ket{\psi_{1}^{R} }= - \ket{\psi_{1}^{R} }$ and $\Upsilon^{L}_{\rm nH} \Upsilon^{R}_{\rm nH} \ket{\psi_{1}^{L} }= - \ket{\psi_{1}^{L} }$ to get the third line of the above relations. Eqs.~(\ref{eq:LtR}, \ref{eq:tLR}) does not hold unless the set of eigenvectors in  $(\Upsilon^{R}_{\rm nH} \ket{\psi^{L}_{1}},\ket{\psi_1^{R}} )$  and in $(\Upsilon^{L}_{\rm nH} \ket{\psi^{R}_{1}}, \ket{\psi_1^{L}})$ are orthogonal, i.e., $\ave{\psi^{L}_{1} |\tilde{\psi}_{1}^{R}  }=\ave{{\psi}_{1}^{R}|  \tilde{\psi}^{L}_{1}   }=0$. 
Hence, I conclude that the presence of a pair of non-Hermitian anti-unitary symmetry $(\Upsilon^{R}, \Upsilon^{L})$ with $\Upsilon^{R} \cdot \Upsilon^{L}=-1$ protects nondefective degeneracies in non-Hermitian systems.  $\blacksquare$

To exemplify our findings, I now explore the stabilization of nondefective degeneracies by non-Hermitian symmetries on a 2D square lattice.

\begin{figure}[t!]
\includegraphics[width=0.75\columnwidth]{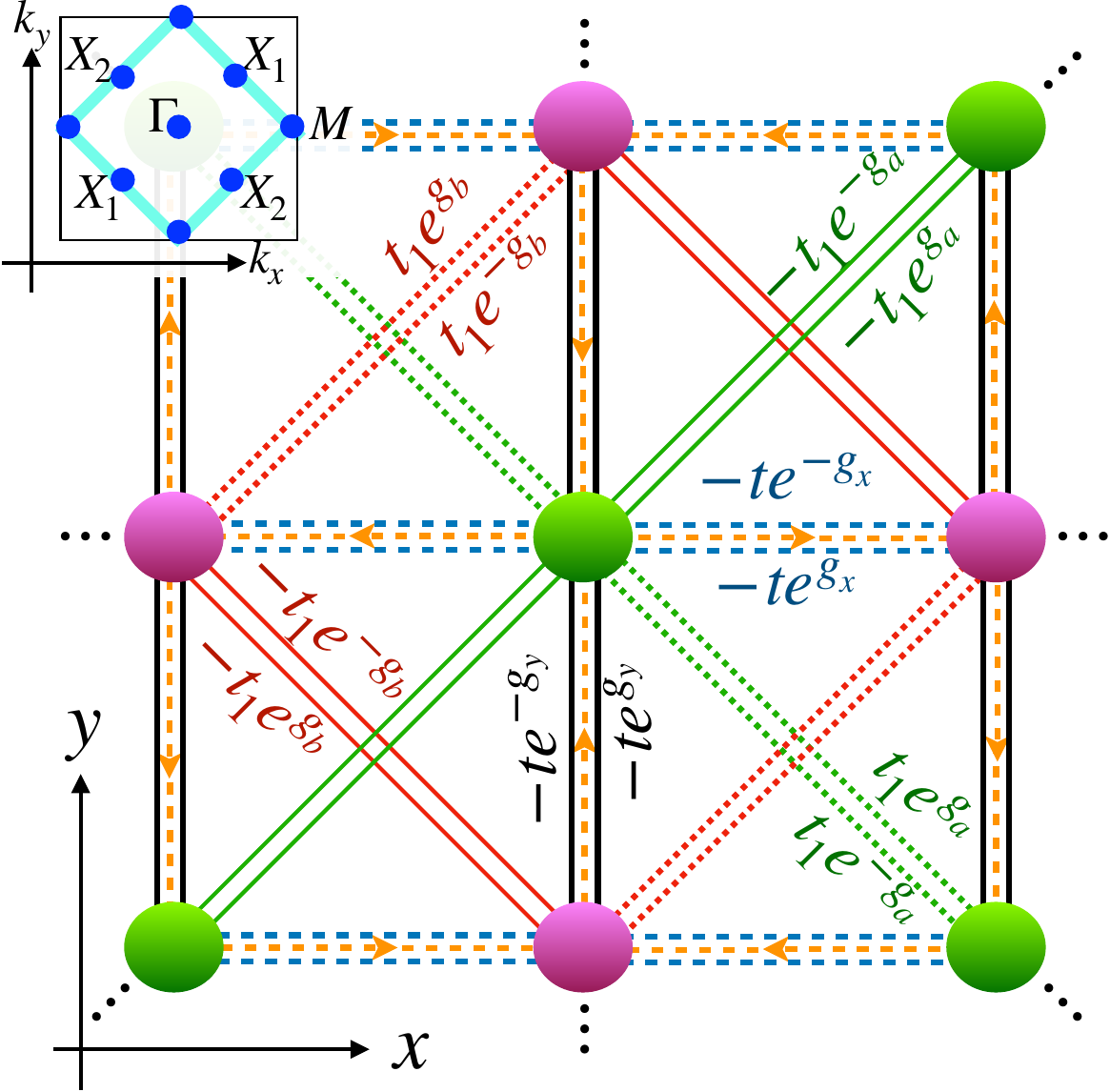}
\caption{ Illustration of the tight-binding model on the square lattice with $A$~(green) and $B$~(magenta) sublattices. Various nearest neighbor and next nearest neighbor hopping amplitudes are shown in solid and dashed lines. The Peierls phase along $x$ or $y$ directions are shown with orange dashed line arrows. The inset displays the first Brillouin zone delineated by the cyan solid line. The high symmetry points, shown in blue, are at $(k_{x},k_{y})=(0,0)$ at the $\Gamma$ point, $(\pi,0)$ at the $M$ point, $\{ (\pi/2,\pi/2),  (-\pi/2,-\pi/2)\}$ at the $X_{1}$ points, and $\{ (\pi/2,-\pi/2),  (-\pi/2,\pi/2) \}$ at the $X_{2}$ points.
\label{fig:lattice_BZI}}
\end{figure}

\paragraph{\bf 2D Tight-binding model.---}
I consider a non-Hermitian bipartite model on a square lattice with staggered potentials, non-reciprocal hopping terms, and nonzero Peierls phase factors for nearest-neighbor hopping amplitudes, schematically shown in Fig.~\ref{fig:lattice_BZI}. Our model Hamiltonian, which is the non-Hermitian generalization of the model in Ref.~\cite{Hou2013}, reads
\begin{align}
{\cal H} =&  {\cal H}_{0} +  {\cal H}_{1} +  {\cal H}_{2} ,
\\
    {\cal H}_{0} =&-t \Big[ e^{-\i \gamma} e^{-g_{x}} a^{\dagger}_{i} b_{i + \hat{x}} + e^{-\i \gamma} e^{-g_{x}} a^{\dagger}_{i} b_{i - \hat{x}}
    \nonumber \\
    &
    +e^{\i \gamma} e^{-g_{y}} a^{\dagger}_{i} b_{i + \hat{y}} + e^{\i \gamma} e^{-g_{y}} a^{\dagger}_{i} b_{i -\hat{y}}
        \nonumber \\
    &
   + e^{\i \gamma} e^{g_{x}} b^{\dagger}_{i+\hat{x}} a_{i} + e^{\i \gamma} e^{g_{x}} b^{\dagger}_{i -\hat{x}} a_{i}
    \nonumber \\
    &
    +e^{-\i \gamma} e^{g_{y}} b^{\dagger}_{i+\hat{y}} a_{i} + e^{-\i \gamma } e^{g_{y}} b^{\dagger}_{i-\hat{y}} a_{i}
    \Big]
    \label{eq:H0}
    ,\\
    {\cal H}_{1} =&- t_{1} \Big[
    e^{-g_{a}} a^{\dagger}_{i} a_{i + \hat{x} +\hat{y}} 
    -  e^{-g_{a}} a^{\dagger}_{i} a_{i + \hat{x} -\hat{y}}
        \nonumber \\
    &
        -e^{-g_{b}} b^{\dagger}_{i} b_{i + \hat{x}+\hat{y}} 
        +  e^{-g_{b}} b^{\dagger}_{i} b_{i +\hat{x} -\hat{y}}
        \nonumber \\
    &
    +e^{g_{a}} a^{\dagger}_{i + \hat{x} +\hat{y}} a_{i}  
    -  e^{g_{a}} a^{\dagger}_{i + \hat{x} -\hat{y}} a_{i} 
        \nonumber \\
    &
        -e^{g_{b}} b^{\dagger}_{i + \hat{x}+\hat{y}} b_{i}  
        +  e^{g_{b}} b^{\dagger}_{i +\hat{x} -\hat{y}} b_{i} 
        \Big],
         \label{eq:H1}
        \\
        {\cal H}_{2} =&v \sum_{i\in A}a^{\dagger}_{i} a_{i}
        -v \sum_{i\in B} b^{\dagger}_{i} b_{i},
         \label{eq:H2}
\end{align}
where $a^{\dagger}_{i} (b^{\dagger}_{i})$ creates an electron at site $i$ in sublattice $A~(B)$ and $\hat{x}~(\hat{y})$ stands for unit vectors along the $x~(y)$ direction. Here $t$ is the nearest-neighbor hopping amplitude, $t_{1}$ denotes the diagonal hopping amplitude, $v$ sets the staggered onsite potential, $\gamma$ is the Peierls phase and $(g_{x},g_{y},g_{a}, g_{b})$ account nonreciprocity. 

The above quadratic Hamiltonian in the momentum space using $F^{\intercal}= (   a_{\bk} , \, b_{\bk})$ casts
 \begin{align}
 \HH(\bk) =& 
F^{\dagger}
 h_{\bk}
F
 , \text{ with }
 h_{\bk}=
\begin{pmatrix}
    h_{11} & h_{12} \\
    h_{21} & h_{22}
\end{pmatrix},
\label{eq:Hk}
  \end{align}
  where $h_{11}=  -t_1 (-e^{-g_a-\i (k_x-k_y)}-e^{g_a+\i (k_x-k_y)}+e^{-g_a-\i (k_x+k_y)}+e^{g_a+\i (k_x+k_y)})+ \i \mu_a+v $,  $h_{22}= t_1 (-e^{-g_b-\i (k_x-k_y)}-e^{g_b+\i (k_x-k_y)}+e^{-g_b-\i (k_x+k_y)}+e^{g_b+\i (k_x+k_y)})- \i \mu_b-v$, $h_{12}= -2 t e^{-g_x-\i \gamma } \cos (k_x)-2 t e^{-g_y+\i \gamma } \cos (k_y) $ and $h_{21}=   -2 t e^{g_x+\i \gamma } \cos (k_x)-2 t e^{g_y-\i \gamma } \cos (k_y)$.
   
The spectrum of this Hamiltonian yields $\epsilon_{\pm} = ( \tr[h_{\bk} ] \pm \sqrt{\eta})/2$, with $\eta=\tr[h_{\bk}]^2 - 4 \det[h_{\bk}]$. Here, $\tr[h_{\bk}]$ and $\det[h_{\bk}]$ denote the trace and the determinant of $h_{\bk}$, respectively. The complex-valued $\eta$ is the discriminant of the characteristic polynomial. The solutions of $\eta=0$ are degeneracies of $h_{\bk}$~\cite{Bergholtz2021, Sayyad2022}.

Our model Hamiltonian accommodates various quantum phases, including the linear Weyl and quadratic double-Weyl semimetals as well as the trivial and topological band insulators; see the SM for details~\cite{SuppMat}.
In the following, I focus on exploring various parameter regimes within which nondefective degeneracies find room to emerge.

 \begin{figure}
  \includegraphics[width=0.95\columnwidth]{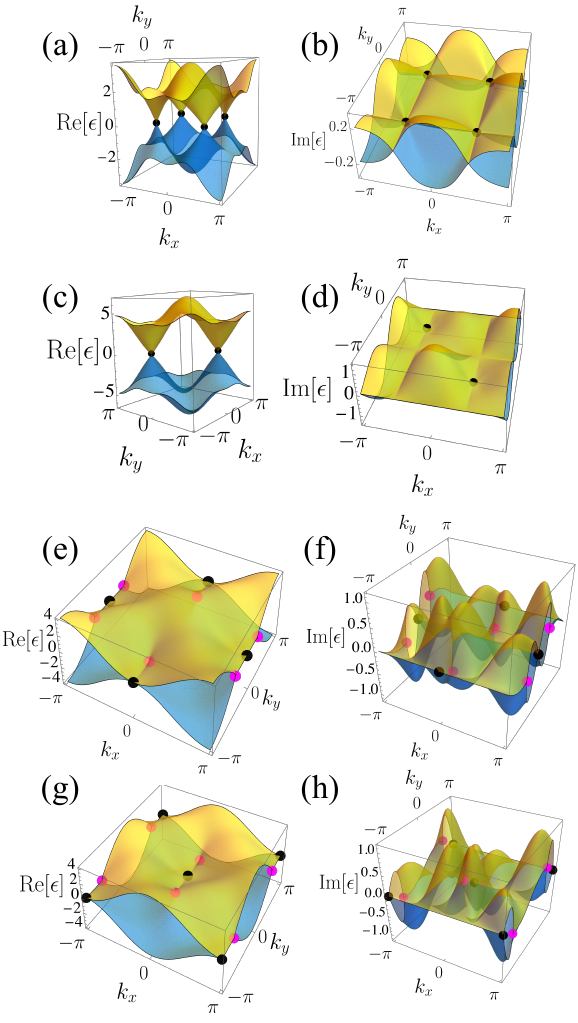}

\caption{ Real~(left panels) and imaginary~(right) parts of the energy dispersion at various parameter regimes.
For each panel, I set $(\gamma, g_{x}, g_{y}, t_{1}/t, g_{a}, g_{b},v/t)=$ $(0.5, 0.5, 0.3, 0.0,0.0,0.0,0.0)\text{~(a,b), }(0.5, 0.0, 0.0, 0.75, 0.5, 0.3,\allowbreak 3.26)\text{~(c,d), }(0.0, 0.0, 0.0, 0.75, 0.5,  0.3, 0.0)\text{~(e,f), }(\pi/2, 0.0, 0.0,\allowbreak 0.75, 0.5, 0.3, 0.0)\text{~(g,h)}$. Magenta points mark the defective EPs, and black points indicate nondefective degeneracies. Line colors are chosen such that the largest~(smallest) values are presented in yellow~(blue).
\label{fig:Spectra_all}}
\end{figure}

 \paragraph{\bf nondefective degeneracies at $t_{1}=g_{a}=g_{b}=v=0$.---}
 The first parameter regime in which nondefective degeneracies arise is at $t_{1}=g_{a}=g_{b}=v=0$. 
Here the model Hamiltonian merely comprises nearest-neighbor hopping terms, i.e., $\HH=\HH_{0}$; see the SM~\cite{SuppMat}. Setting $0< \gamma < \pi/2$ in $\HH$ leads to observing nondefective degeneracies in the band structure of our system. Fig.~\ref{fig:Spectra_all} depicts the real~(a) and imaginary~(b) parts of the spectrum at $\gamma =0.5$, $g_{x}=0.5$ and $g_{y}=0.3$. The nondefective degeneracies, indicated by black points, appear at the $X$ points at $(k_{x},k_{y})=(\pm \pi/2,\pm \pi/2)$ shown in the inset of Fig.~\ref{fig:lattice_BZI} when $\cos(k_{x})=\cos(k_{y})=0$.

The composite symmetry, which guarantees the occurrence of nondefective degeneracies, reads $\Upsilon_{R}= \sigma_{x} {\cal K} {\cal W}_{g} {\cal T}_{\hat{x}} $ and $\Upsilon_{L}= \sigma_{x} {\cal K} {\cal W}_{-g} {\cal T}_{\hat{x}} $ with $\Upsilon_{R} \cdot \Upsilon_{L}={\cal T}_{2\hat{x}}  $. Here ${\cal T}_{\hat{x}}$ is the translational symmetry along the $x$ axis, and ${\cal W}_{g}$ imposes the Wick's rotation on $g_{x}$ and $g_{y}$ such that ${\cal W}_{g} g_{x/y} {\cal W}_{g}^{-1}=\i g_{x/y}$~\footnote{This is analog to the Wick's rotation on the time variable from the Euclidean space into the Minkowski space.}. I note that ${\cal K} {\cal W}_{g}$ is equivalent to ${\cal T}_{+}$ which is the transpose operator used in introducing the time-reversal symmetry, also known as TRS$^\dagger$~\cite{ Kawabata2019b, Kawabata2019, Okuma2020}. Equivalently, instead of ${\cal W}_{g}$, I can introduce an operator~($M_{g}$) which changes the sign of nonHermiticty parameters such that $M^{-1}_{g} g M_{g}=-g$. The non-Hermitian composite symmetry can be readily reduced to the Hermitian Hidden symmetry $\Upsilon_{\rm H}= \sigma_{x} {\cal K}{\cal T}_{\hat{x}} $~\cite{Hou2013} after imposing ${\cal W}_{g}=1$ in the Hermitian limit.  

Considering the right wave-function of the system at the $X $ points as $\ket{\psi^{R}_{X}}$ and acting the $\Upsilon_{R} \cdot \Upsilon_{L}$ on this wave-function gives $\Upsilon_{R} \cdot \Upsilon_{L} \ket{\psi^{R}_{X}} = {\cal T}_{2\hat{x}}  \ket{\psi^{R}_{X}} = e^{- 2 \i x_{X}} \ket{\psi^{R}_{X}} =- \ket{\psi^{R}_{X}} $, where I set the $x$ component of the $X$ points as $x_{X}=\pm \pi/2$. This emphasizes that $\Upsilon_{R} \cdot \Upsilon_{L} $ at nondefective degeneracies is nonunity.

  \paragraph{\bf nondefective degeneracies at nonzero $(t_{1}, g_a,g_b,v) $.---}

 Switching on $\HH_{1}$ or $\HH_2$ violates the $\Upsilon$ symmetry and lifts the degeneracy at (at least two) $X$ points; see the SM for details~\cite{SuppMat}.
Here nonvanishing $\HH_{1,2}$ gives rise to an effective mass term opening a gap at one or both pairs of $X_{1,2}$ nodal points; see the inset of Fig.~\ref{fig:lattice_BZI}. When $v_{ 1}= - 2 t_{1} ( \cosh(g_a) + \cosh(g_b))$ the gap closes at $X_1$ and hence, the system respects $\Upsilon$ at this point. Similarly, the nondefective degeneracy at $X_{2}$ is retrieved when $v_{2}=  2 t_{1} ( \cosh(g_a) + \cosh(g_b))$. I present an example of this situation in Fig.~\ref{fig:Spectra_all}(c,d). Notably, the observed gapless phases at $v_{ 1}$ and $v_{ 2}$ delineate the phase boundaries between two insulating phases, namely, a band insulator and a topological insulator~\cite{SuppMat}.

 \paragraph{\bf nondefective degeneracies at $v=g_{x}=g_{y}=0$ and $\gamma \in \{ 0, \pi/2\}$.---}

 I now study systems at $v=g_{x}=g_{y}=0$ such that the total Hamiltonian consists of ${\cal H} = {\cal H}_{0} + {\cal H}_{1}$. 
 Considering $0< \gamma < \pi/2$ the spectrum of this system is gapped. To be precise, it possesses finite gaps in its real component and has gapless behavior in its imaginary part. However, when $\gamma\in \{ 0 , \pi/2 \}$, the system  hosts only defective EPs with nonzero $(g_{x},g_{y})$ while it exhibits both defective and nondefective degeneracies when $g_{x}=g_{y}=0$; see the SM for more details~\cite{SuppMat}.

Fig.~\ref{fig:Spectra_all}(e,f) presents the spectrum at $\gamma=0$ displaying both defective~(magenta points) and nondefective~(black points) degeneracies around the $M$ points when $g_{x}=g_{y}=0$, $g_{a}=0.5$ and $g_{b}=0.3$. I witness defective degeneracies that result in the bifurcation of the real and imaginary parts of the spectra. I also mark nondefective degeneracies with black points in these figures. The symmetry which protects the nondefective degeneracies reads $\Upsilon'_{R} = \sigma_{x} {\cal K} {\cal W}_{{\cal R }_{g_{a} = g_{b}}}{\cal R}_{x} T_{\hat{x}}$, where  ${\cal R }_{x}$  performs the mirror reflection operation along the $x$ axis such that ${\cal R}_{x}( k_{x}, k_{y}) {\cal R}^{-1}_{x} = ( k_{x}, -k_{y})$. Here, the Wick's rotation acts as ${\cal W}^{-1}_{{\cal R }_{g_{a} = g_{b}}} g_{a} {\cal W}_{{\cal R }_{g_{a} = g_{b}}} = \i g_{b} $ and ${\cal W}^{-1}_{{\cal R }_{g_{a} = g_{b}}} g_{b} {\cal W}_{{\cal R }_{g_{a} = g_{b}}} = \i g_{a} $. 
Defining the right eigenvector at the $M$ points as $\ket{\psi^{R}_{M}}$, I find $\Upsilon'_{R}\cdot \Upsilon'_{L} \ket{\psi^{R}}
=- {\cal T}_{2 \hat{x}} \ket{\psi_{M}^{R}} = -\ket{\psi_{M}^{R}} $, where I have used $\sigma_{x} {\cal R}_{x}= - {\cal R}_{x} \sigma_{x}$. In the Hermitian limit, $\Upsilon'$ acts similar to the Hidden symmetry introduced in Ref.~\cite{Hou2013}.
Notably, the asymptotic behavior of the Hamiltonian close to nondefective degeneracies is quadratic in momenta resulting in the non-Hermitian generalization of quadratic double-Weyl semimetals~\cite{Sun2011, Huang2016, Luo2019, He2020}.

Setting $\gamma=\pi/2$ and $g_{a}= g_{b}=0$, the system again exhibits both defective~(magenta points) and nondefective~(black points) exceptional points, but in this case in the vicinity of the $\Gamma$ points, shown in Fig.~\ref{fig:Spectra_all}(g,h). The system in this parameter regime is invariant under the composite symmetry operator given by $\Upsilon''_{R} = (e^{2 \i \gamma})^{i_{y}} (e^{-2 \i \gamma})^{i_{x}}\sigma_{x} {\cal K} {\cal W}_{{\cal R}_{g_{a} =g_{b}}} {\cal R}_{x} T_{\hat{x}}$, where $(i_{x}, i_{y})$ denote the real-space coordinate of site $i$. I note that $T_{\hat{x}} i_{x} T_{\hat{x}} = (i_{x} + \hat{x}) T_{2\hat{x}}$ and the mirror reflection symmetry with respect to the $x$ axis  results in ${\cal R}_{x} i_{y} {\cal R}_{x} = -i_{y}$. I thereby obtain $\Upsilon''_{R}\cdot \Upsilon''_{L}=-1$ at the $\Gamma$ point. 
Similar to systems at $\gamma=0$, the Hamiltonian close to the nondefective degeneracies is asymptotically quadratic in momenta.

\paragraph{\bf Conclusion.---}
I have proved a theorem stating that anti-unitary symmetries protect all nondefective twofold degeneracies in non-Hermitian systems. I have further exemplified the derivations by exploring the protection of nondefective degeneracies in a 2D tight-binding model with nonreciprocal hopping amplitudes. I have demonstrated that the anti-unitary symmetries constitute various discrete and spatial operations combined with Wick's rotations in the nonreciprocal parameter space. These findings establish the new path to the symmetry-protection of non-Hermitian degeneracies beyond the convention of discrete~\cite{Yoshida2019, Rui2019, Delplace2021, Rivero2021, Sayyad2022, Sayyad2022b} or point-group~\cite{Cui2022} symmetries for defective EPs. Further works should elaborate on the role of the composite symmetries in stabilizing nondefective degeneracies in many-body non-Hermitian systems beyond the commonly studied discrete symmetries~\cite{Takasu2020, Xue2020, Varma2021, Rausch2021, Yoshida2022}. It is also intriguing to explore the robustness of symmetry-protected nondefective degeneracies under small symmetry-breaking perturbations in interacting non-Hermitian systems.

\paragraph{\bf Acknowledgment.---}
I acknowledge the helpful discussion with Jose L. Lado.
I also thank the Galileo Galilei Institute for Theoretical Physics for hospitality during the completion of this work.

\setcounter{secnumdepth}{5}
\renewcommand{\theparagraph}{\bf \thesubsubsection.\arabic{paragraph}}

\renewcommand{\thefigure}{S\arabic{figure}}
\setcounter{figure}{0}

\appendix

\section{Conservation of symmetry operators }

In the following, I present the relations whose satisfaction ensures the conservation of an operator in time. I derive these relations for Hermitian and non-Hermitian systems.

\subsection{Hermitian systems}
A Hermitian system is described by $H_{\rm H} |\psi_{n} \rangle = \epsilon_{n} | \psi_{n} \rangle$, where $H_{\rm H}$ is a generic Hermitian Hamiltonian with eigenvalue $\epsilon_{n}$ and eigenvector $|\psi_{n} \rangle $. For this Hamiltonian, the operator~$\cal O$ is a symmetry operator if it is conserved in time. This quest is translated into
\begin{align}
\partial_{t} \ave{ {\cal O}(t) } &=\ave{\psi_{n} (t) | \HH_{\rm H}^{\dagger } {\cal O} - {\cal O} \HH_{\rm H} |\psi_{n} (t)  }=0,
\label{eq:Ot}
\end{align}
where I used $\ket{\psi_{n} (t)} = \exp(-\i \HH_{\rm H} t ) \ket{\psi_{n} (0)}$.
The Hermiticity of $\HH_{\rm H}$ then simplifies the above relation as
\begin{align}
\HH_{\rm H}^{\dagger } {\cal O} - {\cal O} \HH_{\rm H}&=0 ,\label{eq:Hermrel} \\
  [\HH_{\rm H}, {\cal O}]&=0.\label{eq:Hermrel2}
\end{align}
Hence, ${\cal O}$ is a conserved quantity in Hermitian systems if $ [\HH_{\rm H}, {\cal O}]=0$.

\subsection{Non-Hermitian systems}
A non-Hermitian system is described by 
\begin{align}
\HH_{\rm nH} | \psi_{n}^{R} \rangle &= \epsilon_{n} |\psi_{n}^{R} \rangle,
\qquad \HH_{\rm nH}^{\dagger} | \psi_{n}^{L} \rangle = \epsilon_{n}^{*} |\psi_{n}^{L} \rangle,
\end{align}
where $\HH_{\rm nH}$ is a non-Hermitian Hamiltonian and $\{(\epsilon_{n}, |\psi^{R}_{n} \rangle, |\psi_{n}^{L} \rangle )\}$ describes the non-Hermitian eigensystem with $\ave{\psi_{n}^{L} | \psi_{n}^{R}}=\delta_{mn}$ and $\sum_{n} |\psi_{n}^{R} \rangle \langle \psi_{n}^{L} |=1$.

Based on the biorthogonality of the eigenvectors, I define the biorthogonal operator ${\cal O}\equiv ({\cal O}^{R}, {\cal O}^{L})$. This operator is conserved in time when its elements satisfy
\begin{align}
\begin{cases}
\partial_{t} \ave{{\cal O}^{R}(t)}^{RR} &=\ave{\psi^{R} (t) | {\cal H}^{\dagger}_{\rm eff} {\cal O}^{R} - {\cal O}^{R} {\cal H}_{\rm eff}  |\psi^{R} (t)  }=0
,\\
 \partial_{t}\ave{{\cal O}^{L}(t)}^{LL} &=\ave{\psi^{L} (t) | {\cal H}_{\rm eff}  {\cal O}^{L} - {\cal O}^{L} {\cal H}^{\dagger }_{\rm eff} |\psi^{L} (t)  }=0,
\end{cases}
\label{eq:OtLR}
\end{align}
where $\ave{A}^{RR(LL)}$ is the shortened notation for the expectation value $\ave{\psi^{R(L)} | A | \psi^{R(L)}}$.
Here I used $\ket{\psi^{R}(t)} = \exp(-\i \HH_{\rm nH} t) \ket{\psi^{R} (0)}$ and $\ket{\psi^{L}(t)} = \exp(-\i \HH^{\dagger}_{\rm nH} t) \ket{\psi^{L} (0)}$.

Thereby, to keep this set of symmetry operators stationary in time, the pair of operators should fulfill
\begin{align}
    \begin{cases}
        \HH_{\rm nH}^{\dagger} {\cal O}^{R} - {\cal O}^{R} \HH_{\rm nH} =0
        ,\\
        \HH_{\rm nH} {\cal O}^{L} - {\cal O}^{L} \HH^{\dagger}_{\rm nH}=0.
    \end{cases}
\end{align}
I have employed this set of relations in the main text to prove the theorem.

\section{Various phases and degeneracies in our tight-binding model}
In the main text, I present the nondefective degeneracies which may emerge in our tight-binding model. Here I provide further details on other degeneracies and various phases in our model.

 \subsection{Non-Hermitian linear Weyl semimetal at $t_{1}=g_{a}=g_{b}=v=0$}
\begin{figure*}
  \includegraphics[width=0.75\textwidth]{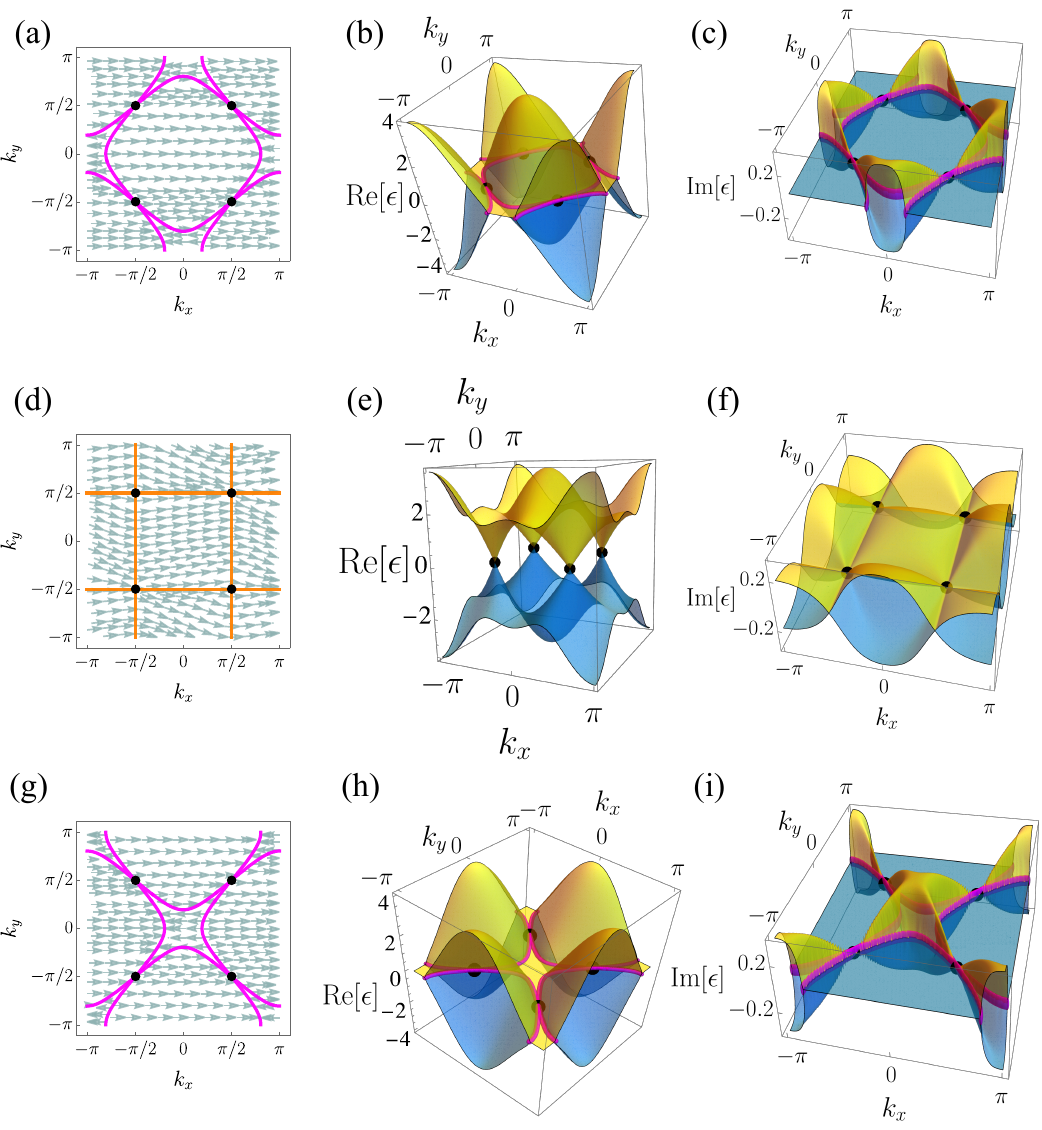}

\caption{ Stream plot of the vector field $(\Re[\eta], \Im[\eta])$~(a,d,g). Real~(b,e,h) and imaginary~(c,f,i) parts of the energy dispersion in Eq.~\eqref{eq:en_weyl}. Parameters are set to $t_{1}/t=g_{a}=g_{b}=v/t=0$, $g_{x}=0.5$, $g_{y}=0.3$ and $\gamma=$ $0.00$~(a,b,c), $0.75$~(d,e,f), $\pi/2$~(g,h,i). Black points depict nondefective degeneracies. Magenta~(Orange) lines display $\Re[\eta]=0~(\Im[\eta]=0)$.
\label{fig:Spectra_weyl}}
\end{figure*}

Setting $t_{1}=g_{a}=g_{b}=v=0$ in Eq.~\eqref{eq:Hk}, the Hamiltonian casts
 \begin{align}
 \HH(\bk) =& 
 \begin{pmatrix}
     a^{\dagger}_{\bk} &
     b^{\dagger}_{\bk}
     \end{pmatrix}
 h_{\bk}
 \begin{pmatrix}
     a_{\bk} \\
     b_{\bk}
 \end{pmatrix}
 ,\\
=&
 \left[
 -2 t e^{-g_x-\i \gamma } \cos (k_x)-2 t e^{-g_y+\i \gamma } \cos (k_y) 
 \right]
 a^{\dagger}_{k} b_{k}
   \nonumber \\
 +&\left[
  -2 t e^{g_x+\i \gamma } \cos (k_x)-2 t e^{g_y-\i \gamma } \cos (k_y)
 \right]
  b^{\dagger}_{k} a_{k}
 .\label{eq:Hmod1k}
\end{align}
The dispersion relation of the system then reads
\begin{align}
    \epsilon_{\pm} =  \pm
    \sqrt{
    f e^{ \phi} +f e^{-\phi} +2  (\cos (2 k_{x})+\cos (2 k_{y})+2)
    },
    \label{eq:en_weyl}
\end{align}
with $\phi=2 \i \gamma +g_{x}-g_{y}$ and $f=4 \cos (k_{x}) \cos (k_{y}) $. To identify degenerate points in the spectrum, one can use the discriminant of the two-band systems given by $\eta=    f e^{ \phi} +f e^{-\phi} +2  (\cos (2 k_{x})+\cos (2 k_{y})+2)$. I present the streamlines of $(\Re[\eta], \Im[\eta])$ in Fig.~\ref{fig:Spectra_weyl}(a,d,g).

Fig.~\ref{fig:Spectra_weyl} presents the real and imaginary parts of the spectrum for $\gamma \in \{ 0, 0.75, \pi/2\}$, $g_{x}=0.5$ and $g_{y}=0.3$.
Based on these results, I can identify three possibilities. 

i)~The first situation happens when $|\epsilon_{\pm}|^2$ is fully real when the Peierls phase vanishes, i.e., at $\gamma=n \pi$ with $n \in \Z$.
I illustrate the streamlines of the vector field $(\Re[\eta], \Im[\eta])$ by teal arrows in Fig.~\ref{fig:Spectra_weyl}(a). Evidently, the orientation of the flow lines reverses at defective EPs when they hit $\Re[\eta]=0$ curves. 
In this case, along the defective exceptional curves, both real and imaginary parts of the spectrum are zero; see magenta lines in panels (b,c). The intersections between these exceptional curves with $\Re[\eta]=0$ mark nondefective degeneracies located at $X$ points and are shown by black points in (a,b,c). As a result, the $X$ points are continuously connected to defective EPs when $\gamma=n \pi$ with $n \in \Z$.
I also witness no changes in the flow direction at nondefective degeenracies $X$ points; see panel (a). This is not unexpected as an even number of EPs which each reverses the stream flow, intersect at the $X$ points, and hence, no alternation of the streamlines should be observed~\cite{Yang2021}.

ii)~The second situation is similar to the previously discussed case i). The only difference is that, in this situation, the coexistence of defective and nondefective degeneracies, appears when $\gamma=n \pi/2$ with odd $n$ values, see Fig.~\ref{fig:Spectra_weyl}(g, h, i).

iii)~The third possibility occurs when $|\epsilon_{\pm}|^2$ is complex-valued with $0<\gamma< \pi/2$. While here, the imaginary part of the spectrum exhibits i-Fermi states, along which $\Im[\epsilon_{\pm}]=0$, the real part of the dispersion relation vanishes merely at nondefective degenracies the $X$ points. Note that these degeneracies are isolated, and they are no longer the termination points of i-Fermi arcs due to the absence of r-Fermi arcs, along which $\Re[\epsilon_{\pm}]=0$~\cite{Yang2021}.

As it is evident from Fig.~\ref{fig:Spectra_weyl}, the nondefective degeneracies at the $X$ points are robust against any changes of $(\gamma, g_{x}, g_{y})$ values. 
This is because our model respects a composite symmetry discussed in the main text.

The asymptotic behavior of the Hamiltonian close to the nondefective degeneracies reads $h^{\rm lin}_{\bk}= \bolds{d} \cdot \bolds{\sigma}$ where $d_{x}= -s[2 t (k_{y} \cosh (g_{y}-i \gamma )-k_{x} \cosh (g_{x}+i \gamma )) ]$, $d_{y}=2 \i  t s (k_{y} \sinh (g_{y}-i \gamma )-k_{x} \sinh (g_{x}+i \gamma ))$ and $d_{z}=0$ with $s=+1(-1)$ at $X_{1}(X_{2})$.
The energy dispersion relation then yields
\begin{align}
    \epsilon^{\rm lin}_{X} &=\mp 2 t  \sqrt{2s k_{x} k_{y} \cosh (2 \i \gamma +g_{x}-g_{y})+k_{x}^2+k_{y}^2},
\end{align}
and the associated eigenvectors read
\begin{align}
    \ket{\psi^{\rm lin}_{X}}=
    \begin{pmatrix}
    \pm \frac{\sqrt{e^{2 \i \gamma } \left(2 s k_{x} k_{y} \cosh (2 i \gamma +g_{x}-g_{y})+k_{x}^2+k_{y}^2\right)}}{e^{g_{y}} k_{y}+k_{x} e^{g_{x}+2 \i \gamma }}
    \\
    1
    \end{pmatrix}.
\end{align}
Moreover, $ h^{\rm lin}_{\bk}$ can be rewritten as a linear Weyl Hamiltonian $h^{\rm lin}_{\bk}= M_{ij} p^{i} \sigma^{j}$ where $M$ is a matrix with complex-valued elements. This Hamiltonian resembles the Hermitian linear Weyl model~\cite{Hou2013}. For this reason, I dub the system in this parameter regime "the non-Hermitian linear Weyl semimetals."

 \begin{figure*}
\includegraphics[width=0.8\textwidth]{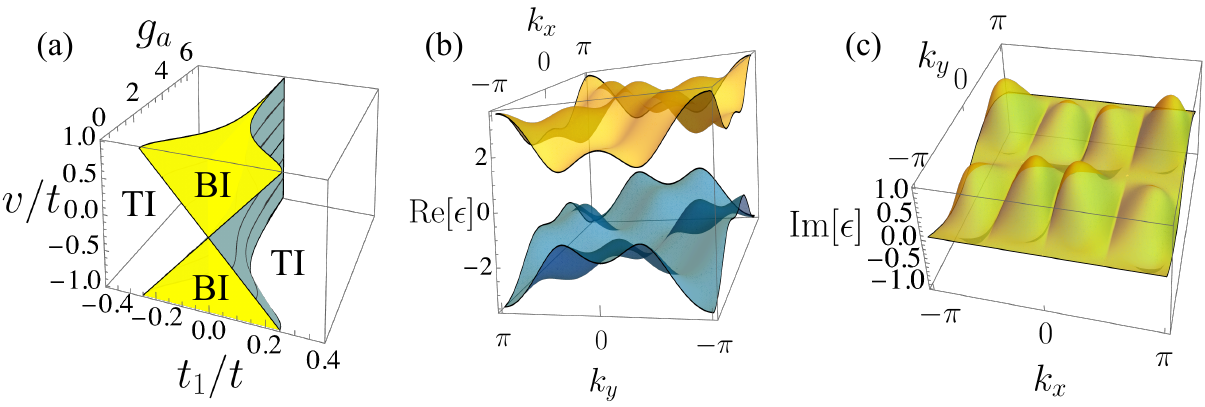}
 
\caption{ (a)~The phase diagram of the system comprises the band insulating~(BI), yellow regions, and topological insulating~(TI) phases at $g_{a}=g_{b}$ and with nonzero $t_{1}$ and $v$. The teal surfaces in (a) separating BI and TI phases depict $v_{1}$ and $v_{2}$ given in Eqs.~(\ref{eq:v1}, \ref{eq:v2}). 
Real~(b) and imaginary~(c) parts of the energy dispersion of $h_{\bk}$ at $t_{1}/t=0.75$, $g_{a}=0.5$, $g_{b}=0.3$ $v/t=g_{x}=g_{y}=0.0$ and $\gamma=0.5$. 
\label{fig:Spect_QAHE_a}}
\end{figure*}

\subsection{Non-Hermitian trivial and topological insulators by keeping either $(t_{1}, g_a,g_b) $ or $v$ nonzero }\label{sec:HallSM}

After turning on $\HH_{1}$ or $\HH_2$, the Hamiltonian $h_{\bk}$ in Eq.~\eqref{eq:Hmod1k} casts $h_{\bk}= \bolds{d}.\bolds{\sigma}$ where $d_{0}=   -4 \i t_1 \sin (k_y) \sinh \left({g_a/2}-{g_b/2}\right) \cosh \left({g_a/2}+{g_b/2}+\i k_x\right)$, $d_{x}=  -2 t \cos (k_x) \cosh (g_x+\i \gamma )-2 t \cos (k_y) \cosh (g_y-\i \gamma )$, $d_{y}=   2 i t \cos (k_x) \sinh (g_x+\i \gamma )+2 i t \cos (k_y) \sinh (g_y-i \gamma )$ and 
 \begin{align}
     d_{z} &=     2 t_1 \left( \cosh (g_a) +\cosh (g_b) \right) \sin (k_x) \sin (k_y)
      \nonumber \\
     &
     -2 \i t_1 \left(  \sinh (g_a) + \sinh (g_b) \right) \cos (k_x) \sin (k_y)
+v.
 \end{align}
Here nonvanishing $d_{z}$ gives rise to an effective mass term, which lifts the degeneracy at, at least two of the $X$ points. The gap closure occurs when
\begin{align}
    v_{1} &= - 2 t_{1} ( \cosh(g_a) + \cosh(g_b)) ,
    \label{eq:v1}\\
    v_{2} &=  2 t_{1} ( \cosh(g_a) + \cosh(g_b)).
    \label{eq:v2}
\end{align}
When $v=v_{1}$, the gap closes at $X_1$. Hence, the system respects the composite symmetry $\Upsilon$ at this point; see the main text. Similarly, the nondefective degeneracy at $X_{2}$ is retrieved when $v=v_{2}$; see also Fig.~\ref{fig:Spectra_all}(c,d) in the main text.

Fig.~\ref{fig:Spect_QAHE_a}(a) displays the phase diagram of our system where the phase boundaries $v_{1,2}$~(teal surfaces) separate the non-Hermitian band insulator~(yellow region) from the non-Hermitian topological insulator. I set $g_{a}=g_{b}$ in Fig.~\ref{fig:Spect_QAHE_a}(a). The complex-valued band structure for the topological insulator with a nonzero gap is exemplified in Fig.~\ref{fig:Spect_QAHE_a}(b,c).

 \begin{figure*}

 \includegraphics[width=0.8\textwidth]{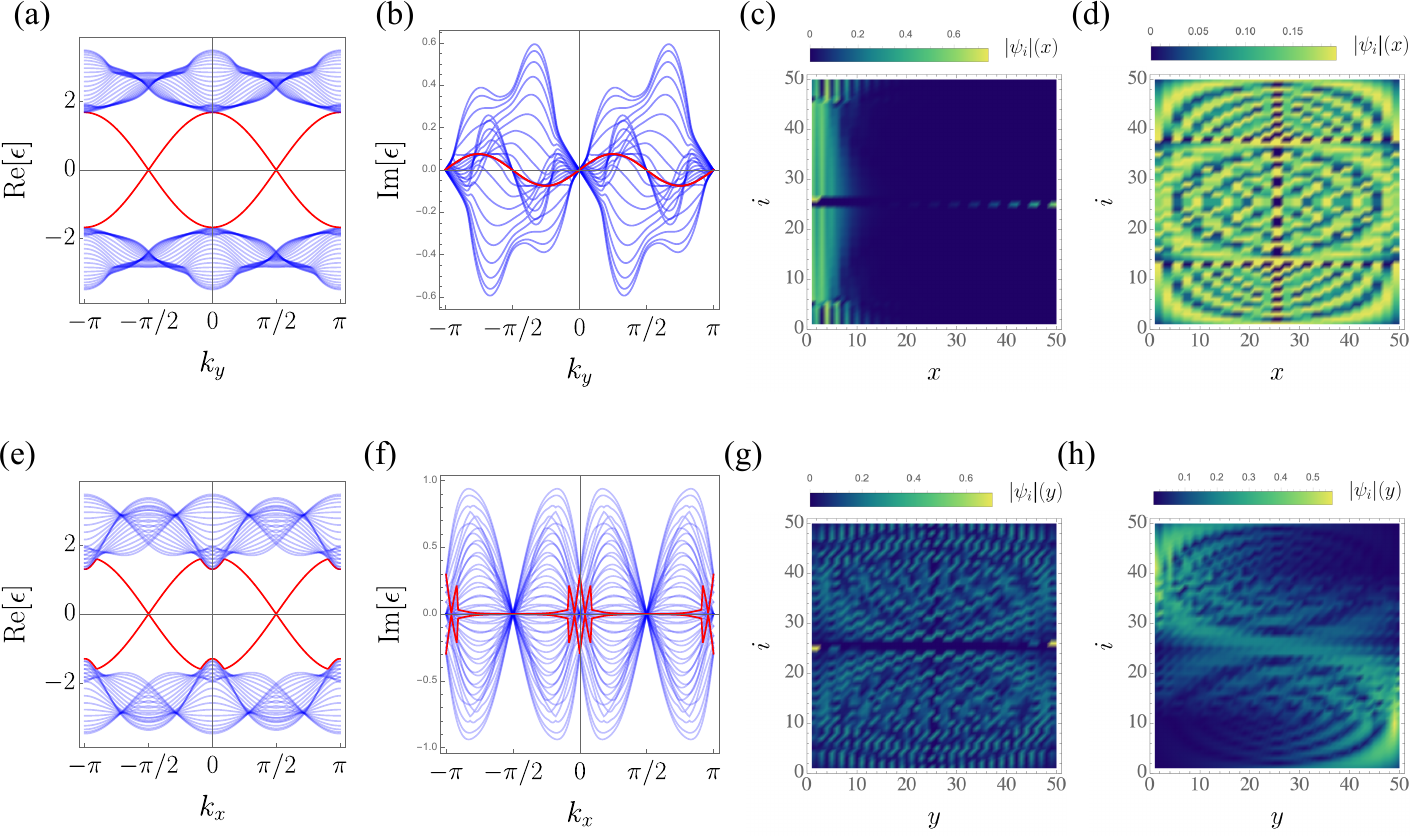}

\caption{ 
Real~(a) and imaginary~(b) parts of the spectrum with open~(periodic) boundary conditions along the x~($k_{y}$)-axis. The absolute values of eigenvectors $|\psi_{i}|$ associated with eigenvalue $\epsilon_{i}$, sorted ascendingly with respect to $\Re[\epsilon_{i}]$, for the system presented in (a,b) at $k_{y}=\pi/2$~(c) and $0$~(d).
(e,f) Similar to (a,b) but with open~(periodic) boundary conditions along the y~($k_{x}$)-axis. The absolute values of eigenvectors for the system are presented in (e,f) at $k_{x}=\pi/2$~(g) and $0$~(h). All plots are obtained at $t_{1}/t=0.75$, $g_{a}=0.5$, $g_{b}=0.3$ $v/t=g_{x}=g_{y}=0.0$ and $\gamma=0.5$.
\label{fig:Spect_QAHE_b}}
\end{figure*}

By imposing open boundary conditions along $x$ or $y$ axis, I can identify chiral edge modes shown in red in Fig.~\ref{fig:Spect_QAHE_b}(a,b) and (e,f), respectively.
These chiral edge modes have finite lifetime~(imaginary parts) for momenta deep inside the gap when the system is merely periodic along the $y$ axis. However, when periodicity is respected only along the $x$ axis, the chiral edge modes possess zero imaginary parts inside the gap region away from the bulk states~(blue curves).
The chiral edge modes in both cases are localized at opposite boundaries of the system, as can be seen from the absolute values of eigenvectors, shown in (c,g), at $i=30,31$ associated with $\epsilon_{i}=0.0$. I further observe the skin effect, the localization of bulk modes at boundaries, e.g., in panel (c, h), due to the reciprocity~(nonzero $(g_{a}, g_{b})$). The observed skin effect in panel (h) is also known as the $Z_{2}$ skin effect~\cite{Okuma2020} protected by the time-reversal symmetry. I note that the absence of the skin effect at $k_{y}=0$ in panel (d) and at $k_{x}=\pi/2$ in panel (g) is due to the zero imaginary parts of all modes resulting in delocalizing the bulk states, similar to Hermitian systems.

 \subsection{Non-Hermitian quadratic double Weyl semimetals at $v=0$ and $\gamma \in \{ 0, \pi/2\}$}

  \begin{figure*}
     \includegraphics[width=0.8\textwidth]{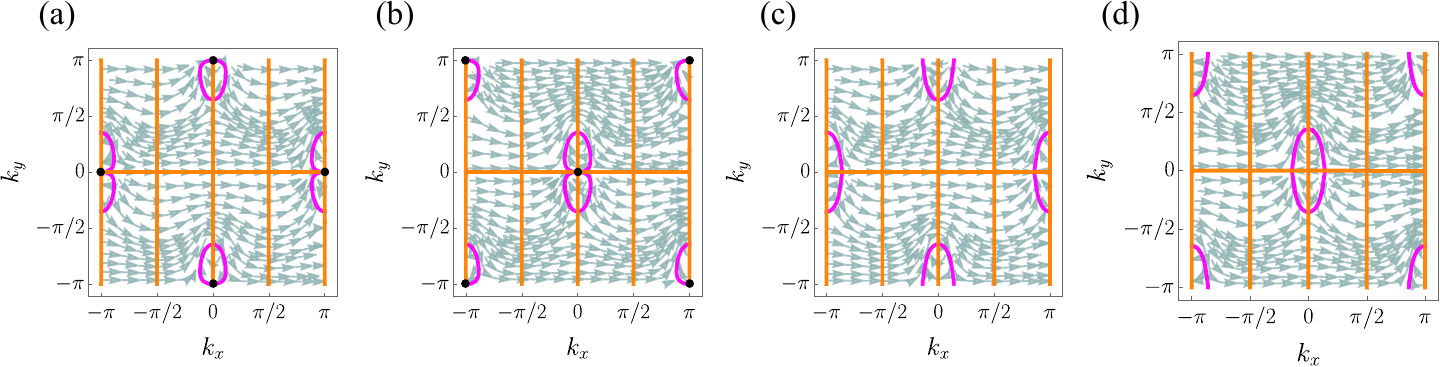} 
\caption{ 
The stream plot of the vector field $(\Re[\eta], \Im[\eta])$ depicted by teal arrows, I present $\Re[\eta]=0$ in magenta solid line and $\Im[\eta]=0$ in orange. Black points indicate nondefective degeneracies. The intersecting points between magenta and orange lines locate defective EPs. 
All plots are obtained at $t_{1}/t=0.75$, $g_{a}=0.5$, $g_{b}=0.3$ $v/t=0.0$. I set $g_{x}=g_{y}=0.0$ and $\gamma=0.0$~in (a), $g_{x}=g_{y}=0.0$ and $\pi/2$ in (b) ,
$g_{x}=0.2$, $g_{y}=0.1$ and $\gamma=0.0$ in (c) and $g_{x}=0.2$, $g_{y}=0.1$ and $\gamma=\pi/2$ in (d).
\label{fig:Spect_2piflux_0}}
\end{figure*}

  \begin{figure*}
 \includegraphics[width=0.8\textwidth]{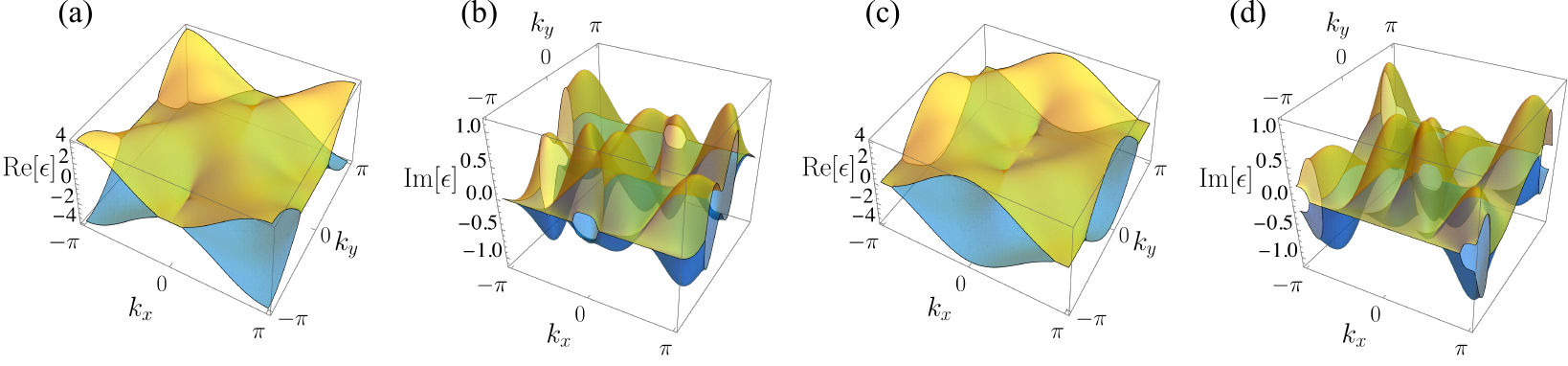}

\caption{ 
Real~(a,c) and imaginary~(b,d) components of the band structure for systems at $t_{1}/t=0.75$, $g_{a}=0.5$, $g_{b}=0.3$ $v/t=0.0$, $g_{x}=0.2$, $g_{y}=0.1$ and $\gamma=0.0$~(a,b) and $\gamma=\pi/2$~(c,d).
\label{fig:Spect_2piflux_00}}
\end{figure*}

Imposing $v=0$ turns off ${\cal H}_{2}$ and subsequently, the Hamiltonian casts $h_{\bk} = d_{0} \id + d_{x} \sigma_{x} + d_{z} \sigma_{z}$ where
\begin{align}
d_{0}=& -4 \i t_{1} \sin (k_{y}) \sinh \left(\frac{  g_{a}-g_{b}}{2} \right) \cosh \left( \frac{ g_{a}+g_{b}+2 \i k_{x} }{2} \right)
,\\
    d_{x} =& 2 \i t [\cos (k_{x}) \sinh (g_{x}+\i \gamma )+\cos (k_{y}) \sinh (g_{y}-\i \gamma )] ,\\
    d_{z} =& 2 t_{1} \sin (k_{y}) [\sin (k_{x}) (\cosh (g_{a})+\cosh (g_{b}))
    \nonumber \\
    &\qquad \quad  -\i \cos (k_{x}) (\sinh (g_{a})+\sinh (g_{b}))].
\end{align}
The associated dispersion relations then reads
$
    \epsilon_{\pm}= d_{0} \pm \sqrt{ \eta}
$ with $\eta=d_{x}^2 + d_{z}^2 $. I present the stream plot of $(\Re[\eta], \Im[\eta])$ in Fig.~\ref{fig:Spect_2piflux_0} at $t_{1}/t=0.75$, $g_{a}=0.5$, $g_{b}=0.3$, $g_{x}=g_{y}=0$, and $\gamma=0$~(a) and $\pi/2$~(b). I also plot Fig.~\ref{fig:Spect_2piflux_0}(c,d) with $t_{1}/t=0.75$, $g_{a}=0.5$, $g_{b}=0.3$, $g_{x}=0.2$, $g_{y}=0.1$ and $\gamma=0$~(c) and $\pi/2$~(d). Fig.~\ref{fig:Spect_2piflux_0} further displays $\Re[\eta]=0$ in magenta and $\Im[\eta]=0$ in orange lines. The intersection of these lines identifies defective EPs in our systems. When even numbers of these lines cross, I detect nondefective degeneracies marked in black points in (a,b). These nondefective degeneracies locate at the $M$ points when $\gamma=0$, and they appear at the $\Gamma$ points at $\gamma = \pi/2$; see also the discussion in the main text. I emphasize that these nondefective degeneracies disappear when nonreciprocal nearest neighbor hopping parameters, namely $(g_{x},g_{y})$, are nonzero, as shown in (c,d). 
The energy band associated with parameters in Fig.~\ref{fig:Spect_2piflux_0}(c,d) are shown in Fig.~\ref{fig:Spect_2piflux_00}.
Unsurprisingly, transitioning from one gapless phase to the other is through a gapped phase with $0< \gamma<\pi/2$.

  \begin{figure*}
\includegraphics[width=0.8\textwidth]{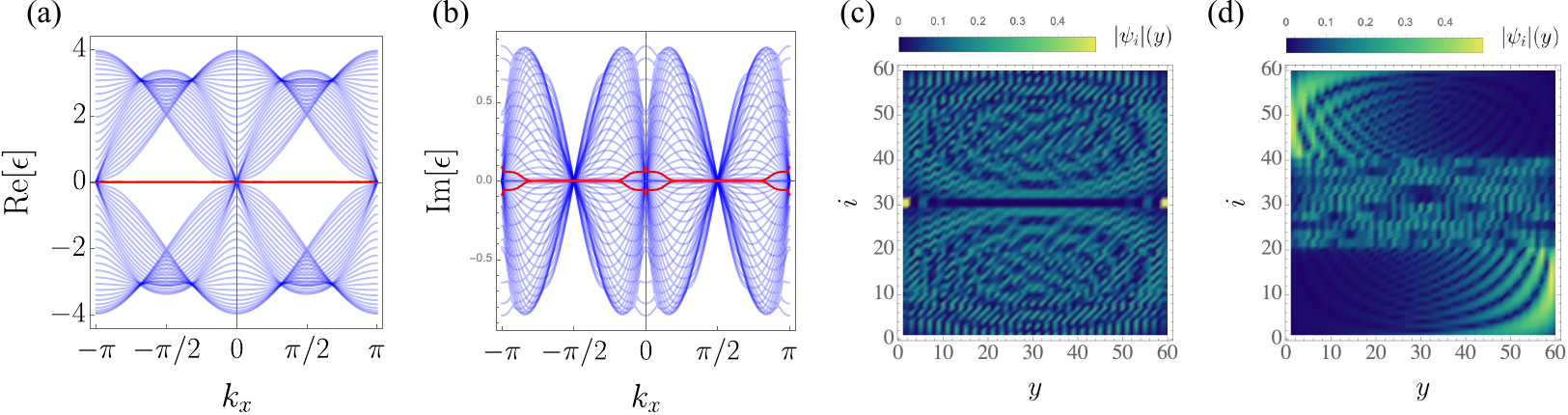}
\caption{ 
The real~(a) and imaginary~(b) parts of the spectrum with open~(periodic) boundary conditions along the x~($k_{y}$)-axis. 
The absolute value of eigenvectors for the system presented in (a,b) at $k_{x}=\pi/2$~(c) and $0$~(d). 
All plots are obtained at $t_{1}/t=0.75$, $g_{a}=0.5$, $g_{b}=0.3$ $v/t=g_{x}=g_{y}=0.0$ and $\gamma=0$.
\label{fig:Spect_2piflux_a0}}
\end{figure*}

 \begin{figure*}
\includegraphics[width=0.8\textwidth]{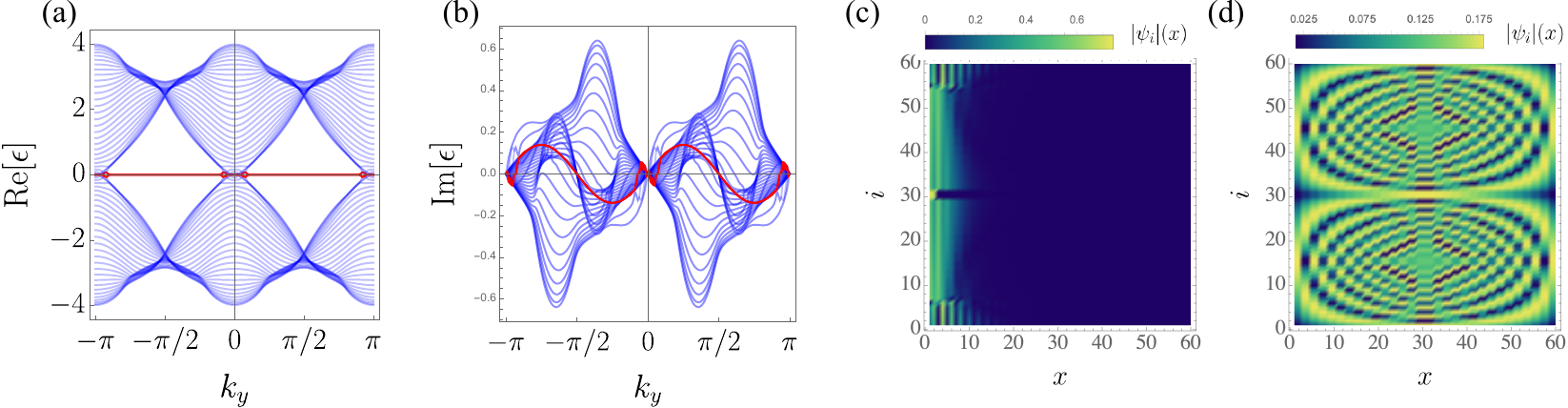}

\caption{ 
The same as Fig.~\ref{fig:Spect_2piflux_a0} but with open~(periodic) boundary conditions along the x~($k_{y}$)-axis. 
\label{fig:Spect_2piflux_a1}}
\end{figure*}

  \begin{figure*}
\includegraphics[width=0.8\textwidth]{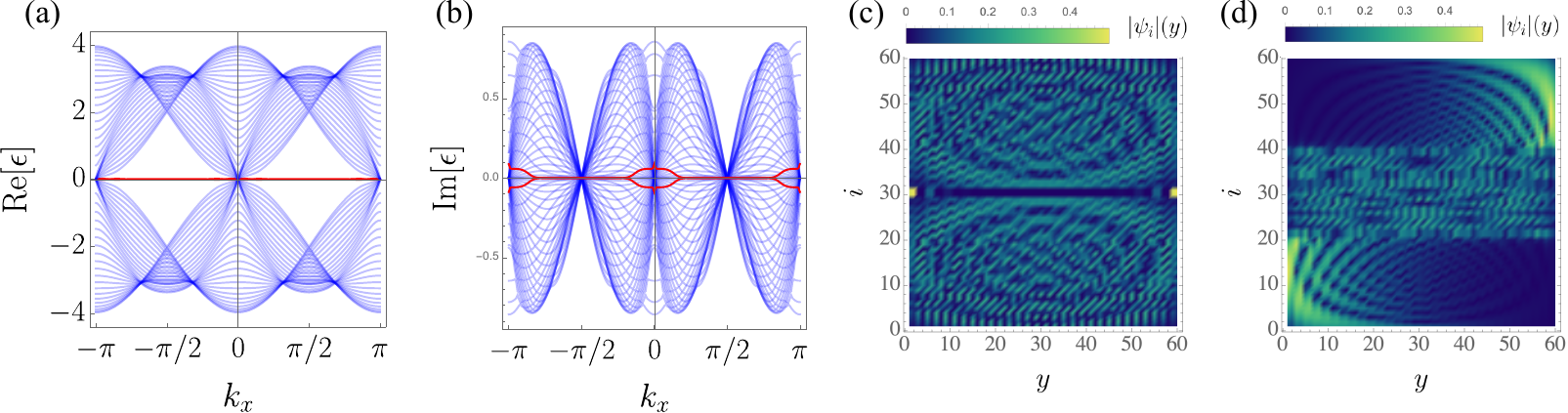}

\caption{ 
The same as Fig.~\ref{fig:Spect_2piflux_a0} but the absolute value of eigenvectors for the system presented in (a,b) at $k_{y}=\pi/2$~(c) and $0$~(d). For all panels I set $t_{1}/t=0.75$, $g_{a}=0.5$, $g_{b}=0.3$ $v/t=g_{x}=g_{y}=0.0$ and $\gamma=\pi/2$. 
\label{fig:Spect_2piflux_b0}}
\end{figure*}

  \begin{figure*}
\includegraphics[width=0.8\textwidth]{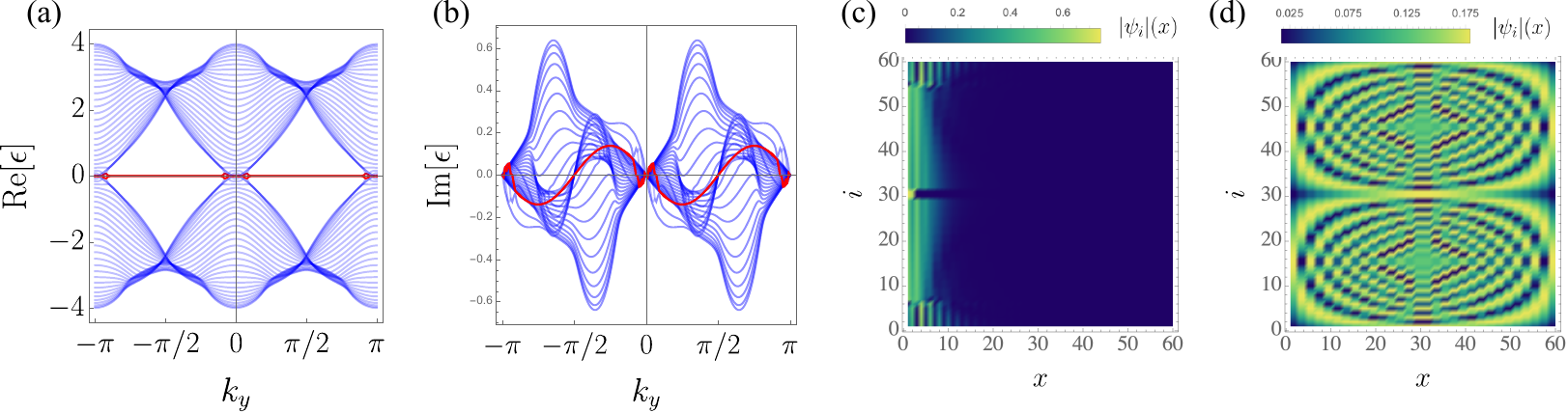}

\caption{ 
The same as Fig.~\ref{fig:Spect_2piflux_a1} but at
$t_{1}/t=0.75$, $g_{a}=0.5$, $g_{b}=0.3$ $v/t=g_{x}=g_{y}=0.0$ and $\gamma=\pi/2$.
\label{fig:Spect_2piflux_b1}}
\end{figure*}

Around nondefective degeneracies, the Hamiltonian is asymptotically quadratic in momentum as it casts $h_{\bk}^{\rm quad} = \bolds{d}^{\rm quad}. \bolds{\sigma}$ where
\begin{align}
    d_{0}^{\rm quad} =& 
    2 p_y t_1 [-\i (\sinh (g_a)-\sinh (g_b))
        \nonumber \\
     \qquad \qquad \quad 
     &
    +p_x \cosh (g_a)-p_x \cosh (g_b)]
    ,\\
    d_{x}^{\rm quad} =&
    t \left[ \left(p_x^2-2\right) \cosh (g_x)+\left(p_y^2-2\right) \cosh (g_y)\right]
        ,\\
    d_{y}^{\rm quad} =&
    -\i t \left[ \left(p_x^2-2\right) \sinh (g_x)+\left(p_y^2-2\right) \sinh (g_y)\right]
        ,\\
    d_{z}^{\rm quad} =&
    2 p_y t_1 [p_x (\cosh (g_a)+\cosh (g_b))
    \nonumber \\
     \qquad \qquad \quad 
     &
    -\i (\sinh (g_a)+\sinh (g_b))]
    ,
\end{align}
where $\bp= \bk - M$, $\gamma=0$ and nondefective degeneracies are located at the $M$ points. When $\gamma=\pi/2$ and nondefective degeneracies reside at the $\Gamma$ points, different components of $h_{\bp}^{\rm quad}$ with $\bp= \bk - \Gamma$ yield
\begin{align}
     d_{0}^{\rm quad} =& 
     2 p_y t_1 [-\i (\sinh (g_a)-\sinh (g_b))
     \nonumber \\
     \qquad \qquad \quad & +p_x \cosh (g_a)-p_x \cosh (g_b)]
         ,\\
    d_{x}^{\rm quad} =&
    \i t \left[ \left(p_x^2-2\right) \sinh (g_x)-\left(p_y^2-2\right) \sinh (g_y)\right]
        ,\\
    d_{y}^{\rm quad} =&
    t \left[ \left(p_x^2-2\right) \cosh (g_x)-\left(p_y^2-2\right) \cosh (g_y)\right]
        ,\\
    d_{z}^{\rm quad} =&
    2 p_y t_1 [p_x (\cosh (g_a)+\cosh (g_b))
    \nonumber \\
     \qquad \qquad \quad  &
    -\i (\sinh (g_a)+\sinh (g_b))]
    .
\end{align}
These systems are the non-Hermitian generalizations of the quadratic double Weyl semimetals~\cite{Sun2011, Huang2016, Luo2019, He2020}.

After understanding our model with the periodic boundary conditions in this parameter regime, I now explore the underlying physics when periodicity along the $x$ or $y$ axis is lifted; See real and imaginary components of the energy spectra in panels (a,b) in Figs.~\ref{fig:Spect_2piflux_a0}, \ref{fig:Spect_2piflux_a1}, \ref{fig:Spect_2piflux_b0} and \ref{fig:Spect_2piflux_b1}. Here Figs.~\ref{fig:Spect_2piflux_a0} and \ref{fig:Spect_2piflux_a1} are plotted at $\gamma=0$ and Figs.~\ref{fig:Spect_2piflux_b0} and \ref{fig:Spect_2piflux_b1} are obtained at $\gamma=\pi/2$. Evidently, these two sets of results are very similar to each other. Both cases exhibit boundary modes with nearly zero real and imaginary parts when periodicity along the $k_y$ axis is relaxed. On the contrary, the boundary modes possess finite imaginary parts when one enforces the open boundary conditions along the $x$ axis. 
The behavior of eigenvectors in these parameter regimes is similar to those discussed in Sec.~\ref{sec:HallSM}. The only major difference is how eigenvectors behave at two zero eigen-energies~($\epsilon_{i}=0.0$) with $i=30,31$ in panels (c) in Figs.~\ref{fig:Spect_2piflux_a0}, \ref{fig:Spect_2piflux_a1}, \ref{fig:Spect_2piflux_b0}, and \ref{fig:Spect_2piflux_b1}. For these two eigenvalues, in contrast to the results of Sec.~\ref{sec:HallSM}, I realize that the associated eigenvectors coalesce. Hence, I conclude that these systems with open boundary conditions along the $y$ and $x$ axes experience defective EPs at $k_{x}=\pi/2$ and $k_{y}=\pi/2$, respectively.
I also observe the localization of the bulk eigenvectors, i.e., the skin effect, around $x=0$; see Figs.~\ref{fig:Spect_2piflux_a1} and \ref{fig:Spect_2piflux_b1}.
However, the bulk eigenvectors are localized at the left/right ends of the system along the $y$-axis when periodicity along $k_y$ is relaxed.

\bibliography{bibfile}

\end{document}